\documentclass[%
 reprint,
superscriptaddress,
amsmath,amssymb,
aps,
pra,
]{revtex4-2}

\usepackage{graphicx}
\usepackage{dcolumn}
\usepackage{bm}
\usepackage{color}
\usepackage{braket}

\begin{document}

\preprint{APS/123-QED}

\title{Quantum uncertainty of optical coherence}

\author{Martti~Hanhisalo}
 \email{martti.hanhisalo@uef.fi}
\affiliation{%
 Center for Photonics Sciences, University of Eastern Finland, P.O. Box 111, FI-80101 Joensuu, Finland
}%

\author{Mohammad~Sajjad~Mirmoosa}
\affiliation{%
 Center for Photonics Sciences, University of Eastern Finland, P.O. Box 111, FI-80101 Joensuu, Finland
}%
\author{Tero~Set\"{a}l\"{a}}%
\affiliation{%
 Center for Photonics Sciences, University of Eastern Finland, P.O. Box 111, FI-80101 Joensuu, Finland
}%
\author{{\L}ukasz~Rudnicki}
\affiliation{%
 Center for Photonics Sciences, University of Eastern Finland, P.O. Box 111, FI-80101 Joensuu, Finland
}%
\affiliation{%
  International Centre for Theory of Quantum Technologies, University of Gda\'nsk,\\ Jana Ba\.zy\'nskiego 1a, 80-308 Gda\'nsk, Poland
 }%
\author{Andreas~Norrman}
\affiliation{%
 Center for Photonics Sciences, University of Eastern Finland, P.O. Box 111, FI-80101 Joensuu, Finland
}%

\date{\today}

\begin{abstract}
Light is known to exhibit quantum uncertainty in terms of its amplitude, phase, and polarization. However, quantum uncertainty related to coherence, which is also a fundamental physical property of light, has not been considered to date. Here, we formulate and explore the concept of quantum optical coherence uncertainty. We focus on the first-order coherence of the simplest possible light field, a purely monochromatic plane wave, which is classically completely stable. Starting from a scalar treatment, we show that the field displays zero coherence uncertainty only for a number state. We then proceed to the vectorial regime and establish that any state leads to coherence fluctuations, governed by a set of uncertainty relations depending on the polarization state and space-time points. Our work thus provides fundamental insights into the quantum character of optical coherence, with potential benefits in applications using highly sensitive interferometric and polarimetric techniques.
\end{abstract}

\maketitle


\section{Introduction}
The uncertainty principle is a primary concept in quantum physics and a major manifestation of quantum complementarity~\cite{Heisenberg,Bohr}. It states that quantum objects admit complementary observables whose values cannot be known simultaneously with arbitrary precision. For light, this uncertainty arises as quantum fluctuations in amplitude and phase~\cite{Agarwal:13}, whose nonclassical features are utilized in quantum-enhanced sensing and metrology including gravitational wave detection~\cite{Aasi:13,Tse:19}, biological measurements~\cite{Bowen:13,Taylor:16}, and even dark-matter probing~\cite{Backes:21}. When the vector nature of light is taken into account, the uncertainty is manifested in the form of quantum polarization fluctuations that exist for any state of light~\cite{Luis:16,Soto:21}. This fact imposes fundamental boundaries for optical polarimetry but opens up also opportunities for quantum-polarimetric techniques~\cite{Goldberg:22} that can outperform classical methods~\cite{Rudnicki:20}.

Coherence is likewise a central notion in optical physics and a fundamental property of light, whose understanding and exploitation is crucial in many different contexts, e.g., interference, diffraction, propagation, spectral distribution, and interactions~\cite{Mandel:95}. Especially the quantum aspects of optical coherence~\cite{Glauber:63} have been in the spotlight in various branches of physics. Besides its wide interest in quantum optics, such as interferometry~\cite{Pleinert:21}, imaging~\cite{Schneider:18}, metrology~\cite{Su:17}, information processing~\cite{Fortsch:13}, and testing quantum foundations~\cite{Deng:19}, it is met in optomechanics~\cite{Cohen:15}, molecular physics~\cite{Rosenberg:22}, quantum electronics~\cite{Brange:15}, and cosmological studies~\cite{Giovannini:11}. Moreover, the interplay of coherence and polarization has been an active research subject in classical optics~\cite{Friberg:16}, triggering also related quantum investigations~\cite{Norrman:17,Norrman:20}. Nevertheless, to the best of our knowledge, the role of quantum uncertainty affiliated with coherence has not been examined before.

In this work, we introduce and study the concept of first-order coherence uncertainty of quantized light fields. We particularly focus on a purely monochromatic plane wave, which is classically totally coherent and displays no fluctuations either in amplitude, phase, or polarization. Firstly, we show in the scalar context that such field exhibits zero coherence uncertainty only for a number state. We then elucidate how for any other state the coherence fluctuations change when the space-time points are varied in terms of a coherence phase space. Secondly, we extend our analysis to the vector-light regime and show that for any state the coherence uncertainty is nonzero due to the inevitable quantum polarization fluctuations. We further establish a set of uncertainty relations for the coherence fluctuations, which depend not only on the polarization state but also on the space-time coordinates. Similarly to the scalar-light treatment, we visualize the space-time dependent coherence fluctuations within the vectorial regime through a phase-space representation. The exact assessment of coherence is thereby generally and fundamentally prohibited by quantum uncertainty. Hence, besides amplitude, phase, and polarization, our work reveals previously uncharted and complementary facets of quantum light uncertainty via optical coherence.

\section{Coherence uncertainty of scalar light}
The first-order correlations of a quantized scalar light field are, at two space-time points $x_1=(\mathbf{r}_1,t_1)$ and $x_2=(\mathbf{r}_2,t_2)$, fully characterized by the coherence function~\cite{Glauber:63}
\begin{equation}
    G(x_1,x_2)=\mathrm{tr}[\hat{\rho}\hat{E}^\dagger(x_1)\hat{E}(x_2)].\label{coherence-function}
\end{equation}
Here, $\hat{E}(x)$ is the positive frequency part of the electric field operator, $\hat{\rho}$ is the density operator specifying the quantum state, $\dagger$ denotes the Hermitian adjoint, and tr stands for the trace in Fock space. When the two points coincide, $G(x,x)$ is the local intensity at $x$ and proportional to the photodetection rate~\cite{Glauber:63}. Thus the respective Hermitian operator $\hat{G}(x,x)=\hat{E}^\dagger(x)\hat{E}(x)$ represents a physical observable. On the other hand, when $x_1\neq x_2$, the coherence function $G(x_1,x_2)$ is generally a complex number and not directly measurable, whereupon the corresponding coherence function operator $\hat{G}(x_1,x_2)=\hat{E}^\dagger(x_1)\hat{E}(x_2)$ is non-Hermitian. However, the coherence function can be determined piecewise by interferometric means. In two-path interferometry, its magnitude and phase are experimentally ascertained from the visibility and the locations, respectively, of the intensity fringes~\cite{Mandel:95}. Alternatively, the coherence function can be characterized by its real and imaginary parts, i.e. $G(x_1,x_2)=G'(x_1,x_2)+\mathrm{i}G''(x_1,x_2)$. The operators associated with these two parts,
\begin{subequations}
    \begin{align}
    \hat{G}'(x_1,x_2)&=\frac{1}{2}[\hat{G}(x_1,x_2)+\hat{G}^\dagger(x_1,x_2)],\label{G'}\\
    \hat{G}''(x_1,x_2)&=\frac{1}{2\mathrm{i}}[\hat{G}(x_1,x_2)-\hat{G}^\dagger(x_1,x_2)],\label{G''}
\end{align}
\end{subequations}
are clearly Hermitian, with their expectations yielding the complete information on the whole coherence function $G(x_1,x_2)$.

Let us consider a monochromatic plane-wave mode
\begin{equation}
\hat{E}(x)=C\hat{a}\mathrm{e}^{\mathrm{i}(\mathbf{k}\cdot\mathbf{r}-\omega t)},\label{scalar-E}
\end{equation}
in which $C$ is a constant, $\hat{a}$ is the annihilation operator, $\mathbf{k}$ is the wave vector, and $\omega$ is the angular frequency. Now the two Hermitian operators in Eqs.~(\ref{G'}) and (\ref{G''}) read
\begin{subequations}
    \begin{align}
    \hat{G}'(x_1,x_2)&=|C|^2\hat{n}\cos\Theta,\label{G'pw}\\
    \hat{G}''(x_1,x_2)&=|C|^2\hat{n}\sin\Theta,\label{G''pw}
\end{align}
\end{subequations}
where $\hat{n}$ is the number operator and $\Theta=\mathbf{k}\cdot(\mathbf{r}_2-\mathbf{r}_1)-\omega(t_2-t_1)$. We immediately observe that the expectations $G'(x_1,x_2)=|C|^2\bar{n}\cos\Theta$ and $G''(x_1,x_2)=|C|^2\bar{n}\sin\Theta$, with $\bar{n}$ being the mean photon number, are exactly $\pi/2$ out of phase. More importantly, from Eqs.~(\ref{G'pw}) and (\ref{G''pw}) we find the coherence uncertainties (standard deviations)
\begin{subequations}
    \begin{align}
    \Delta G'(x_1,x_2)&=|C|^2\Delta n|\cos\Theta|,\label{varG'}\\
    \Delta G''(x_1,x_2)&=|C|^2\Delta n|\sin\Theta|,\label{varG''}
\end{align}
\end{subequations}
where $\Delta n$ is the uncertainty in the photon number.

Equations~(\ref{varG'}) and (\ref{varG''}) show that the uncertainties in the real and imaginary parts of the coherence function can simultaneously vanish only when $\Delta n=0$, i.e., for number states, which are the eigenstates of $\hat{G}'(x_1,x_2)$ and $\hat{G}''(x_1,x_2)$. As these states represent the purest form of sub-Poissonian light, the vanishing of both $\Delta G'(x_1,x_2)$ and $\Delta G''(x_1,x_2)$ has no classical correspondence. Any other state ($\Delta n>0$) exhibits coherence uncertainty at all space-time points due to the $\pi/2$ phase difference between $\Delta G'(x_1,x_2)$ and $\Delta G''(x_1,x_2)$, even if the field is a perfectly monochromatic plane wave. Accordingly, except the case of a number state, the complete determination of the coherence properties of the field cannot be achieved, not even in principle, with arbitrary precision. The distribution of the uncertainty between the real and imaginary parts of the coherence function is specified through $\Theta$ by the space-time points $x_1$ and $x_2$. On the other hand, whereas the coherence uncertainty alternates between $\Delta G'(x_1,x_2)$ and $\Delta G''(x_1,x_2)$ when changing the space-time points, the squared sum of these fluctuations is independent of space and time:
\begin{equation}
[\Delta G'(x_1,x_2)]^2+[\Delta G''(x_1,x_2)]^2=|C|^4(\Delta n)^2.\label{variance-sum-G}
\end{equation}
This behavior is somewhat similar to the energy swinging back and forth between a harmonic oscillator's potential and kinetic energies, while the total energy remains constant.

\begin{figure}
    \centering
    \includegraphics[scale=1]{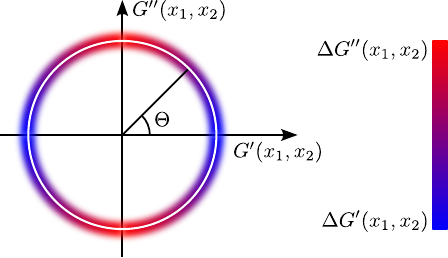}
    \caption{Schematic of the coherence phase space for a quantized monochromatic plane wave in the scalar-light domain. The real part $G'(x_1,x_2)$ and the imaginary part $G''(x_1,x_2)$ of the coherence function (white circle) exhibit respective uncertainties $\Delta G'(x_1,x_2)$ and $\Delta G''(x_1,x_2)$ (red-blue ring). The distribution of the uncertainties is specified by the space-time dependent angle $\Theta$ and highlighted with the color coding.}
    \label{Fig1}
\end{figure}

The above findings can be visualized by introducing the coherence phase space illustrated in Fig.~\ref{Fig1}, where the horizontal and vertical axes correspond to $G'(x_1,x_2)$ and $G''(x_1,x_2)$, respectively. In this representation, the coherence function $G(x_1,x_2)$ is located at a distance of $|G(x_1,x_2)|=|C|^2\bar{n}$ from the origin and at an angle of $\Theta$ with respect to the horizontal axis. As the space-time separation is varied, the coherence function evolves along a circle, surrounded by an uncertainty ring specified by $\Delta G'(x_1,x_2)$ and $\Delta G''(x_1,x_2)$. We stress that $\Theta$ is completely deterministic and, thus, the uncertainty in Fig.~\ref{Fig1} appears only in the radial direction, corresponding to the magnitude $|G(x_1,x_2)|$. According to Eq.~(\ref{variance-sum-G}), the uncertainty of the coherence magnitude is dictated exclusively by the photon-number fluctuations.

Moreover, and more generally, the real and imaginary parts of the coherence function in Eq.~(\ref{coherence-function}) are directly linked to the magnitude $|G(x_1,x_2)|$ and phase $\mathrm{arg}[G(x_1,x_2)]$. As explained below Eq.~(\ref{coherence-function}), these quantities and thus also $G'(x_1,x_2)$ and $G''(x_1,x_2)$ can be obtained experimentally from interferometric measurements. By the same token, since $\Delta G'(x_1,x_2)$ and $\Delta G''(x_1,x_2)$ may in general lead to fluctuations both in the magnitude and phase of $G(x_1,x_2)$, the coherence uncertainty is expected to occur as quantum fluctuations in the visibility and the locations of the interference fringes. This in turn prevents the characterization of the coherence properties with zero uncertainty. Hence, from an operational point of view, such quantum fluctuations may limit high-precision control and measurement of optical coherence~\cite{Divitt:16,Morrill:16,Li:17} as well as the image resolution and signal estimation in coherence-based metrology~\cite{Larson:18,Hradil:19,Hradil:21,Wadood:21,De:21,Liang:21,Ares:21}.


\section{Polarization uncertainty}
In a moment we will generalize the above framework to the vectorial regime, i.e., light whose polarization properties must be taken into account. But first we briefly review some necessary background concepts related to the quantum polarization uncertainty of such vectorial light fields.

Let the column vector $\hat{\mathbf{E}}(x)=[\hat{E}_h(x),\,\hat{E}_v(x)]^\mathrm{T}$ denote a two-component electric field operator (the superscript T stands for the transpose), with $\hat{E}_h(x)$ and $\hat{E}_v(x)$ being the two orthogonal components along some arbitrary horizontal and vertical axes, respectively. The quantum polarization properties of such a light field can be characterized by the four Stokes operators~\cite{Norrman:20}
\begin{equation}    
\hat{S}_n(x)=\hat{\mathbf{E}}^\dagger(x)\boldsymbol{\sigma}_n\hat{\mathbf{E}}(x),\quad n\in\{0,1,2,3\},\label{one-point-general}
\end{equation}
where $\boldsymbol{\sigma}_0$ is the $2\times2$ identity matrix and $\boldsymbol{\sigma}_1,\boldsymbol{\sigma}_2,\boldsymbol{\sigma}_3$ are the Pauli matrices. Contrary to the coherence function operator, the Stokes operators in Eq.~(\ref{one-point-general}) are Hermitian.
The expectation values of Eq.~(\ref{one-point-general}),
\begin{equation}
S_n(x)=\mathrm{tr}[\hat{\rho}\hat{S}_n(x)],\quad n\in\{0,1,2,3\},\label{one-point-expectations}
\end{equation}
correspond to the classical Stokes parameters~\cite{Gil:22}. These four parameters satisfy the constraint $|\mathbf{S}(x)|\leq S_0(x)$, with $\mathbf{S}(x)=[S_1(x),S_2(x),S_3(x)]^\mathrm{T}$, and have the following physical interpretations: $S_0(x)$ is proportional to the (average) total photon number, while $S_1(x)$, $S_2(x)$, and $S_3(x)$ describe the (average) differences between $h$- and $v$-polarized, $+\pi/4$- and $-\pi/4$-linearly polarized, and right- and left-circularly-polarized photons, respectively.

Considering a vectorial monochromatic plane wave
\begin{equation}
    \hat{\mathbf{E}}(x)=C[\hat{a}_h,\,\hat{a}_v]^\mathrm{T}\mathrm{e}^{\mathrm{i}(\mathbf{k}\cdot\mathbf{r}-\omega t)},\label{electric-field}
\end{equation}
in which $\hat{a}_s$ is the annihilation operator of the $s\in\{h,v\}$ polarization mode, we obtain from Eq.~(\ref{one-point-general}) the (space-time independent) Stokes operators~\cite{Luis:16,Soto:21}
\begin{equation}
\begin{gathered}
    \!\!\!\hat{S}_0=|C|^2(\hat{a}_h^\dagger\hat{a}_h+\hat{a}^\dagger_v\hat{a}_v),\ \; \hat{S}_1=|C|^2(\hat{a}_h^\dagger\hat{a}_h-\hat{a}^\dagger_v\hat{a}_v),\\
    \!\!\!\hat{S}_2=|C|^2(\hat{a}_h^\dagger\hat{a}_v+\hat{a}^\dagger_v\hat{a}_h),\ \; \hat{S}_3=\mathrm{i}|C|^2(\hat{a}_v^\dagger\hat{a}_h-\hat{a}^\dagger_h\hat{a}_v).\label{one-point-stokes}
\end{gathered}
\end{equation}
Due to the commutator $[\hat{a}_s,\hat{a}_{s'}^\dagger]=\delta_{ss'}$, with $\delta$ being the Kronecker delta, the Stokes operators in Eq.~(\ref{one-point-stokes}) obey the commutation relations~\cite{Luis:16,Soto:21}
\begin{equation}
    [\hat{S}_0,\hat{\mathbf{S}}]=\mathbf{0},\quad [\hat{S}_j,\hat{S}_k]=2\mathrm{i}|C|^2\epsilon_{jkl}\hat{S}_l,\label{polarization-commutators}
\end{equation}
where $\hat{\mathbf{S}}=[\hat{S}_1,\hat{S}_2,\hat{S}_3]^\mathrm{T}$ and $\epsilon_{jkl}$ is the Levi--Civita tensor. The first relation in Eq.~(\ref{polarization-commutators}) concurs with the classical perception that polarization ($\hat{\mathbf{S}}$) and intensity ($\hat{S}_0$) are separate physical entities: the form of the ellipse (polarization) is independent of its size (intensity). However, in contrast to the classical framework, the second relation in Eq.~(\ref{polarization-commutators}) implies that not even a perfectly monochromatic plane wave can have a definite, nonfluctuating polarization state due to the noncommutative components of $\hat{\mathbf{S}}$. In particular, the latter commutator in Eq.~(\ref{polarization-commutators}) leads to the polarization uncertainty relations~\cite{Luis:16,Soto:21}
\begin{equation}
    \Delta S_j\Delta S_k\geq |C|^2|\epsilon_{jkl}||S_l|,\label{polarization-uncertainty}
\end{equation}
which set the fundamental precision boundaries for the simultaneous assessment of the Stokes parameters $S_1$, $S_2$, and $S_3$.


\section{Coherence uncertainty of vector light}
We now turn to examine coherence uncertainty in the vector-light regime. We start by introducing the coherence matrix operator
\begin{equation}
    \hat{\mathbf{G}}(x_1,x_2)=\hat{\mathbf{E}}^\dagger(x_1)\otimes\hat{\mathbf{E}}(x_2),\label{coherence-matrix-operator}
\end{equation}
where $\otimes$ denotes the outer product and the field operator is generally a three-component vector. The expectation
\begin{equation}
    \mathbf{G}(x_1,x_2)=\mathrm{tr}[\hat{\rho}\hat{\mathbf{G}}(x_1,x_2)] \label{eq1}
\end{equation}
corresponds to the classical $3\times3$ coherence matrix~\cite{Friberg:16} and contains all the information on the first-order space-time coherence of a general three-component field.

Here we focus on a two-component field, characterized by $\hat{\mathbf{E}}(x)=[\hat{E}_h(x),\,\hat{E}_v(x)]^\mathrm{T}$. In this scenario $\hat{\mathbf{G}}(x_1,x_2)$ is a $2\times2$ matrix, formed by the non-Hermitian elements $\hat{G}_{\alpha\beta}(x_1,x_2)=\hat{E}_\alpha^\dagger(x_1)\hat{E}_\beta(x_2)$ with $\alpha,\beta\in\{h,v\}$. Akin to the scalar coherence function in Eq.~(\ref{coherence-function}), each coherence-matrix element $G_{\alpha\beta}(x_1,x_2)=\mathrm{tr}[\hat{\rho}\hat{G}_{\alpha\beta}(x_1,x_2)]$ is hence a complex quantity with a real part $G_{\alpha\beta}'(x_1,x_2)$ and an imaginary part $G_{\alpha\beta}''(x_1,x_2)$. As with Eqs.~(\ref{G'}) and (\ref{G''}), we could then define the respective Hermitian operators $\hat{G}_{\alpha\beta}'(x_1,x_2)$ and $\hat{G}_{\alpha\beta}''(x_1,x_2)$, which would allow us to assess the coherence uncertainty of the field.

However, it is more instructive to employ the coherence Stokes operators~\cite{Norrman:20}
\begin{equation}    \hat{S}_n(x_1,x_2)=\hat{\mathbf{E}}^\dagger(x_1)\boldsymbol{\sigma}_n\hat{\mathbf{E}}(x_2),\quad n\in\{0,1,2,3\},\label{Two-point-general}
\end{equation}
offering an alternative yet equivalent way to represent the first-order coherence of two-component quantized light. The coherence Stokes operators are two-point extensions of the standard polarization (one-point) Stokes operators in Eq.~(\ref{one-point-general}), and it is readily verified that they are related to the $2\times2$ coherence matrix operator in Eq.~(\ref{coherence-matrix-operator}) as
\begin{align}
    \hat{\mathbf{G}}(x_1,x_2)&=\frac{1}{2}\sum_{n=0}^3 \hat{S}_n(x_1,x_2)\boldsymbol{\sigma}_n,\\
    \hat{S}_n(x_1,x_2)&=\mathrm{Tr}[\hat{\mathbf{G}}(x_1,x_2)\boldsymbol{\sigma}_n],
\end{align}
where Tr stands for the matrix trace. Furthermore, their expectation values
\begin{equation}
    S_n(x_1,x_2)=\mathrm{tr}[\hat{\rho}\hat{S}_n(x_1,x_2)],\quad n\in\{0,1,2,3\},\label{Two-point-expectations}
\end{equation}
are the quantum analogs of the classical coherence Stokes parameters~\cite{Ellis:04,Korotkova:05}. Physically, $S_0(x_1,x_2)$ describes the sum of correlations of the $h$ and $v$ components of the electric field at the two space-time points, while $S_1(x_1,x_2)$, $S_2(x_1,x_2)$, and $S_3(x_1,x_2)$ characterize the correlation differences between the $h$- and $v$-polarized, $+\pi/4$- and $-\pi/4$-linearly polarized, and right- and left-circularly-polarized components, respectively~\cite{Tervo:09}.

When $x_1=x_2=x$, the coherence Stokes operators (parameters) in Eq.~(\ref{Two-point-general}) [Eq.~(\ref{Two-point-expectations})] reduce to their Hermitian (real) polarization counterparts in Eq.~(\ref{one-point-general}) [Eq.~(\ref{one-point-expectations})]. But, when $x_1\neq x_2$, the coherence Stokes operators are non-Hermitian, meaning that the generally complex coherence Stokes parameters are not physical observables. Given this fact and motivated by Eqs.~(\ref{G'}) and (\ref{G''}), we therefore introduce the Hermitian operators
\begin{subequations}
    \begin{align}
    \hat{S}_n'(x_1,x_2)&=\frac{1}{2}[\hat{S}_n(x_1,x_2)+\hat{S}_n^\dagger(x_1,x_2)],\label{S'}\\
    \hat{S}_n''(x_1,x_2)&=\frac{1}{2\mathrm{i}}[\hat{S}_n(x_1,x_2)-\hat{S}_n^\dagger(x_1,x_2)],\label{S''}
\end{align}
\end{subequations}
corresponding to the real part $S'_n(x_1,x_2)$ and imaginary part $S''_n(x_1,x_2)$, respectively, of the complex coherence Stokes parameter $S_n(x_1,x_2)=S_n'(x_1,x_2)+\mathrm{i}S_n''(x_1,x_2)$. 
Altogether, these eight two-point quantities contain the complete information on the first-order coherence of a two-component field. This contrasts with the full polarization characterization of the field, which instead requires the knowledge of only four one-point Stokes parameters.

We further remark that the real and imaginary parts of the complex coherence Stokes parameters in Eq.~(\ref{Two-point-expectations}) can equally well be described by the magnitudes $|S_n(x_1,x_2)|$ and phases $\arg[S_n(x_1,x_2)]$. In vector-light interferometry, these magnitudes and phases can be experimentally inferred from the intensity and polarization-state fringes of the corresponding one-point Stokes parameters~\cite{Friberg:16, Turunen:22}. This technique then also allows to determine $S'_n(x_1,x_2)$ and $S''_n(x_1,x_2)$. Akin to the scalar-light case, any possible uncertainty related to $S'_n(x_1,x_2)$ and $S''_n(x_1,x_2)$ should subsequently show up as quantum fluctuations in the fringes of the one-point Stokes parameters. We anticipate such fluctuations to set quantum constraints in both conventional~\cite{Kanseri:09,Kanseri:10,Leppänen:17,Partanen:18} and more recent nanoscattering~\cite{Leppänen:15,Chen:17,Saastamoinen:20} and polarimetric~\cite{Laatikainen:24} schemes for probing vector-light coherence.

Considering the monochromatic plane wave in Eq.~(\ref{electric-field}), from Eqs.~(\ref{Two-point-general}), (\ref{S'}), and (\ref{S''}) we then readily obtain
\begin{subequations}
    \begin{align}
         \hat{S}_n'(x_1,x_2)&=\hat{S}_n\cos\Theta,\label{S'pw}\\
        \hat{S}_n''(x_1,x_2)&=\hat{S}_n\sin\Theta.\label{S''pw}
    \end{align}
\end{subequations}
Equations~(\ref{S'pw}) and (\ref{S''pw}) are the vector-light extensions of Eqs.~(\ref{G'pw}) and (\ref{G''pw}), which connect the Hermitian coherence Stokes operators $\hat{S}_n'(x_1,x_2)$ and $\hat{S}_n''(x_1,x_2)$ to their associated polarization Stokes operators $\hat{S}_n$ in Eq.~(\ref{one-point-stokes}). This link offers interesting insights into the commutation relations of $\hat{S}_n'(x_1,x_2)$ and $\hat{S}_n''(x_1,x_2)$.

Firstly, as regards $\hat{S}_0'(x_1,x_2)$ and $\hat{S}_0''(x_1,x_2)$, we find from Eqs.~(\ref{S'pw}), (\ref{S''pw}), and the first relation in (\ref{polarization-commutators}) that
\begin{subequations}
\begin{equation}
[\hat{S}_0'(x_1,x_2),\hat{S}_0''(x_1,x_2)]=0.\label{commutator0a}
\end{equation}
Likewise, the commutators of $\hat{S}_0'(x_1,x_2)$ and $\hat{S}_0''(x_1,x_2)$ with $\hat{S}_n'(x_1,x_2)$ and $\hat{S}_n''(x_1,x_2)$ for $n\in\{1,2,3\}$ fulfill
\begin{equation}    [\hat{S}_0^\mu(x_1,x_2),\hat{S}_n^\nu(x_1,x_2)]=0,\quad \mu,\nu\in\{\prime,\prime\prime\}.\label{commutator0b}
\end{equation}
\end{subequations}
Accordingly, the operators $\hat{S}_0^\prime(x_1,x_2)$ and $\hat{S}_0^{\prime\prime}(x_1,x_2)$ do not only commute mutually, but also with any of the two Hermitian parts of the other coherence Stokes operators. The latter property shares formal similarity to the first relation in Eq.~(\ref{polarization-commutators}) of the polarization Stokes operators. Physically it means that the simultaneous precise assessment of the correlation sums encoded in $S_0'(x_1,x_2)$ and $S_0''(x_1,x_2)$, and the correlation differences embedded in the real and imaginary parts of the remaining coherence Stokes parameters, is in principle possible.

However, and secondly, in view of Eqs.~(\ref{S'pw}) and (\ref{S''pw}) as well as the latter relation in Eq.~(\ref{polarization-commutators}) the commutators of $\hat{S}_n'(x_1,x_2)$ and $\hat{S}_n''(x_1,x_2)$ for $n\in\{1,2,3\}$ obey
\begin{subequations}
\begin{align}
    [\hat{S}_j'(x_1,x_2),\hat{S}_k'(x_1,x_2)]&=2\mathrm{i}|C|^2\epsilon_{jkl}\hat{S}_l\cos^2\Theta,\label{commutator1}\\
    [\hat{S}_j''(x_1,x_2),\hat{S}_k''(x_1,x_2)]&=2\mathrm{i}|C|^2\epsilon_{jkl}\hat{S}_l\sin^2\Theta,\label{commutator2}\\
    [\hat{S}_j'(x_1,x_2),\hat{S}_k''(x_1,x_2)]&=\mathrm{i}|C|^2\epsilon_{jkl}\hat{S}_l\sin(2\Theta).\label{commutator3}
\end{align}    
\end{subequations}
We infer that these coherence Stokes operators are governed by three cyclic commutation relations, contrary to the single cyclic commutator in Eq.~(\ref{polarization-commutators}) for the polarization Stokes operators. Furthermore, the commutators of the coherence Stokes operators depend both on the polarization Stokes operators and the space-time points via the phase $\Theta$. The different trigonometric functions in Eqs.~(\ref{commutator1})--(\ref{commutator3}) especially signify that the coherence Stokes parameters $S_n'(x_1,x_2)$ and $S_n''(x_1,x_2)$ cannot be determined simultaneously for all $n\in\{1,2,3\}$ with an arbitrary accuracy. For example, considering $j\neq k\neq l$, the commutator in Eq.~(\ref{commutator1}) [Eq.~(\ref{commutator2})] tends to zero merely when $\cos\Theta=0$ ($\sin\Theta=0$). In this case also the commutator in Eq.~(\ref{commutator3}) vanishes but the commutator in Eq.~(\ref{commutator2}) [Eq.~(\ref{commutator1})] is nonzero and reduces to the latter polarization commutation relation in Eq.~(\ref{polarization-commutators}).

\begin{figure}[h]
    \centering
    \includegraphics{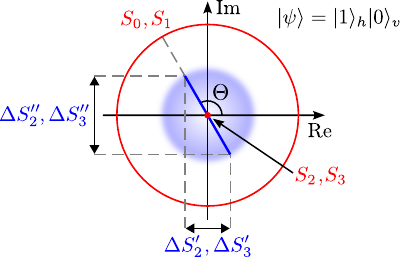}
    \caption{Illustration of the coherence Stokes parameters and their uncertainties (arguments omitted for brevity) for a monochromatic plane wave [Eq.~(\ref{electric-field})] in state $\ket{\psi}=\ket{1}_h\ket{0}_v$. The parameters $S_0(x_1,x_2)=S_1(x_1,x_2)=|C|^2\mathrm{e}^{\mathrm{i}\Theta}$ showing no uncertainties follow the sharp red circle when the phase $\Theta$ evolves. However, the parameters $S_2(x_1,x_2)=S_3(x_1,x_2)=0$ (red dot at the origin) display the fluctuations $\Delta S_2'(x_1,x_2)=\Delta S_3'(x_1,x_2)=|C|^2|\cos\Theta|$ and $\Delta S_2''(x_1,x_2)=\Delta S_3''(x_1,x_2)=|C|^2|\sin\Theta|$ (blue circle) in their real and imaginary parts.}
    \label{Fig2}
\end{figure}

Such coherence uncertainty can be formulated in terms of the standard deviations $\Delta S_n'(x_1,x_2)$ and $\Delta S_n''(x_1,x_2)$. Because any two Hermitian operators $\hat{A}$ and $\hat{B}$ satisfy $\Delta A\Delta B\geq|\braket{[\hat{A},\hat{B}]}|/2$~\cite{Sakurai}, we immediately obtain from Eqs.~(\ref{commutator1})--(\ref{commutator3}) the uncertainty relations
\begin{subequations}
\begin{align}
    \!\!\!\Delta S_j'(x_1,x_2)\Delta S_k'(x_1,x_2)&\geq|C|^2|\epsilon_{jkl}||S_l|\cos^2\Theta,\label{uncertainty-relation1}\\
    \!\!\!\Delta S_j''(x_1,x_2)\Delta S_k''(x_1,x_2)&\geq|C|^2|\epsilon_{jkl}||S_l|\sin^2\Theta,\label{uncertainty-relation2}\\
    \!\!\!\Delta S_j'(x_1,x_2)\Delta S_k''(x_1,x_2)&\geq\frac{|C|^2}{2}|\epsilon_{jkl}||S_l||\sin(2\Theta)|.\label{uncertainty-relation3}
\end{align}    
\end{subequations}
On the other hand, when $\mathbf{S}=\mathbf{0}$ the right-hand sides of Eqs.~(\ref{uncertainty-relation1})--(\ref{uncertainty-relation3}) automatically disappear and the uncertainty relations become trivial. To bypass this triviality, we examine $\Delta S_n'(x_1,x_2)$ and $\Delta S_n''(x_1,x_2)$ individually. From Eqs.~(\ref{S'pw}) and (\ref{S''pw}) we find for any $n\in\{0,1,2,3\}$
\begin{subequations}
    \begin{align}
        \Delta S_n'(x_1,x_2)&=\Delta S_n|\cos\Theta|,\label{varS'}\\
        \Delta S_n''(x_1,x_2)&=\Delta S_n|\sin\Theta|,\label{varS''}
    \end{align}
\end{subequations}
where $\Delta S_n$ is the standard deviation of the associated polarization Stokes parameter. Equations~(\ref{varS'}) and (\ref{varS''}), which together with Eqs.~(\ref{commutator0a})--(\ref{uncertainty-relation3}) form the central result of this work, are the vector-light generalizations of the scalar-light relations in Eqs.~(\ref{varG'}) and (\ref{varG''}). However, while for scalar light the coherence uncertainty becomes strictly zero for a number state, in the vectorial regime this is never possible. The reason for such unavoidable coherence uncertainty for vector light is that the fluctuations of all polarization Stokes parameters cannot vanish for any state~\cite{Luis:16,Soto:21} and because of the $\pi/2$ phase difference between $\Delta S_n'(x_1,x_2)$ and $\Delta S_n''(x_1,x_2)$. Accordingly, in the vector-light framework, even a perfectly monochromatic plane-wave field displays intrinsic coherence uncertainty regardless of the quantum state. Moreover, while the coherence uncertainty oscillates between $\Delta S_n'(x_1,x_2)$ and $\Delta S_n''(x_1,x_2)$ when the space-time separation is varied, their squared sum is space-time independent:
\begin{equation}
    [\Delta S_n'(x_1,x_2)]^2+[\Delta S_n''(x_1,x_2)]^2=(\Delta S_n)^2.\label{variance-sum-S}
\end{equation}
Equation~(\ref{variance-sum-S}) extends the scalar-light relation in Eq.~(\ref{variance-sum-G}) to the vector domain and highlights how the total coherence uncertainty of a vectorial light field is specified by the polarization uncertainty.

As an example, Fig.~\ref{Fig2} provides a phase-space representation of such vector-coherence uncertainty for the state $\ket{\psi}=\ket{1}_h\!\ket{0}_v$. In this situation, the parameters $S_0(x_1,x_2)=S_1(x_1,x_2)=|C|^2\mathrm{e}^{\mathrm{i}\Theta}$ have no uncertainties, but $S_2(x_1,x_2)=S_3(x_1,x_2)=0$ show the $\Theta$-dependent fluctuations $\Delta S_2'(x_1,x_2)=\Delta S_3'(x_1,x_2)=|C|^2|\cos\Theta|$ and $\Delta S_2''(x_1,x_2)=\Delta S_3''(x_1,x_2)=|C|^2|\sin\Theta|$. Hence, even a monochromatic plane wave containing only a single horizontally polarized photon possesses vectorial coherence fluctuations.

\section{Conclusions}
In summary, we have established and examined the notion of quantum coherence uncertainty for scalar light and vectorial light. After introducing general Hermitian operators to characterize the first-order coherence for both light types, we studied in detail the coherence uncertainty of a monochromatic plane-wave field. We first showed that in the scalar framework the quantum fluctuations of the coherence function vanish only in the case of a number state. We also illustrated how for other states the uncertainty alternates between the real and imaginary parts of the coherence function when the space-time points change via a coherence phase space. For vector light, however, we found that no state can lead to zero uncertainty of all coherence Stokes parameters owing to the intrinsic quantum polarization fluctuations. The quantum fluctuations of the real and imaginary parts of the coherence Stokes parameters were further shown to obey a set of uncertainty relations, depending both on the polarization state and on the space-time coordinates. By using a coherence phase-space representation, an example of the space-time dependent quantum fluctuations of the coherence Stokes parameters was also provided. Hence, our work establishes a theoretical foundation for quantum uncertainty of optical coherence and provides a platform for further exploration and exploitation of quantum light correlations in interferometry and polarimetry.

\section*{Acknowledgments}
The authors thank Elvis Pillinen, Jyrki Laatikainen, and A. T. Friberg for fruitful discussions and correspondence. This work was
supported by the Research Council of Finland (Grant
Nos.~354918, 349396, and 346518 as well as PROFI funding
336119).

\end{document}